\documentclass[preprint]{aastex}
\usepackage{epsf}
\usepackage{natbib}
\usepackage{graphicx}
\usepackage{float}
\usepackage{afterpage}
\usepackage{rotating}
 \usepackage{relsize}
\usepackage{amsmath}
\usepackage{epsfig}
\usepackage{enumitem}
\usepackage{bm,bbm,amssymb,color, amsmath, color}
\usepackage{enumerate}
\usepackage{url}
\usepackage[utf8x]{inputenc}

\usepackage{hyperref}

\begin{document}
\title{Producing Distant Planets by Mutual Scattering of Planetary Embryos}
\author{Kedron Silsbee\altaffilmark{1} \& Scott Tremaine\altaffilmark{2}}
\altaffiltext{1}{Department of Astrophysical Sciences, 
Princeton University, Ivy Lane, Princeton, NJ 08544; 
ksilsbee@mpe.mpg.de}
\altaffiltext{2}{Institute for Advanced Study, 
1 Einstein Drive, Princeton, NJ 08540; 
tremaine@ias.edu}

\begin{abstract}
It is likely that multiple bodies with masses between those of Mars and Earth (``planetary embryos") formed in the outer planetesimal disk of the solar system. Some of these were likely scattered by the giant planets into orbits with semi-major axes of hundreds of AU. Mutual torques between these embryos may lift the perihelia of some of them beyond the orbit of Neptune, where they are no longer perturbed by the giant planets so their semi-major axes are frozen in place. We conduct N-body simulations of this process, and its effect on smaller planetesimals in the region of the giant planets and the Kuiper belt. We find that (i) there is a significant possibility that one sub-Earth mass embryo, or possibly more, is still
present in the outer solar system; (ii) the orbit of the surviving embryo(s) typically has perihelion of 40--70 AU, semi-major axis less than 200 AU, and inclination less than 30$^\circ$; (iii) it is likely that any surviving embryos could be detected by current or planned optical surveys or have a significant effect on solar-system ephemerides; (iv) whether or not an embryo has survived to the present day, their dynamical influence earlier in the history of the solar system can explain the properties of the detached disk (defined in this paper as containing objects with perihelia $>$ 38 AU and semi-major axes between 80 and 500 AU).  
 \end{abstract}

\section{Introduction}

\noindent
The standard model for the formation and evolution of comets \citep{Oort50} assumes that comets form in the protoplanetary disk, in the region of the giant planets, and are then excited onto highly eccentric planet-crossing orbits through planetary perturbations.  The comets then undergo a random walk in energy space, taking one ``step" at each perihelion passage when they interact with the giant planets.  This process continues until the aphelion distance is a few times $10^4$ AU, at which point torques from passing stars and the tidal field of the Galaxy are sufficient to increase the perihelion beyond $\sim 38$ AU in less than the energy diffusion time.  Once this occurs, the comet no longer interacts gravitationally with the planetary system and its semi-major axis is frozen in place, apart from a much slower random walk due to perturbations from passing stars.  This model correctly predicts many properties of the Oort comet cloud \citep{Wiegert99, Fernandez05, Fouchard13}, but also incorrectly predicts that there should be no objects with perihelia substantially beyond the orbit of Neptune and aphelia less than a thousand AU, since there is no dynamical pathway to this part of phase space \citep{Duncan87}.  
More specifically, the JPL small-body database\footnote{\url{https://ssd.jpl.nasa.gov/sbdb_query.cgi}} contains 33 objects on orbits with semi-major axes between 80 and 500 AU and perihelia well beyond the orbit of Neptune ($q >$ 38 AU).  Such bodies are often said to belong to the detached disk \citep{Gladman02}.  The best known member of this group is Sedna, a body with perihelion 76 AU, semi-major axis 480 AU, and a radius of about 1,000 km \citep{Brown04, Pal12}.

One model to explain the orbits of these bodies \citep{FB00,Brasser06, Brasser12} holds that the Sun was born in a cluster of stars with density in the range $10^4$--$10^5 M_\odot \, {\rm pc}^{-3}$.  Tides and close stellar encounters from this cluster exert torques on bodies with aphelia of several hundred AU, raising their perihelia and thereby creating the detached disk.  Another possibility is that some of the objects in the detached disk were trapped in mean-motion resonances when Neptune migrated, after which their perihelia were lifted above 40 AU by resonant oscillations \citep{KS16,Nesvorny+16}. Alternatively, \citet{Madigan16} find that collective effects allow a massive disk of planetesimals to increase their inclinations at the expense of their eccentricity, thus producing a population of detached objects.  Several other models for producing the detached disk are reviewed by \citet{Morbidelli04}.

Another idea, first suggested by \citet{Gladman02} but not explored in detail, is that multiple Mars-sized embryos formed interior to the orbit of Neptune, and, like the comets, were excited to semi-major axes of a few hundred AU via interactions with the giant planets.  These embryos exerted torques on one another as well as on smaller bodies and some of their perihelia grew as a result, so their orbits were decoupled from the perturbations of the giant planets and formed a long-lived detached disk.  Here, we investigate this scenario through direct N-body simulations and make predictions for the properties of the detached disk and the masses, numbers, and orbits of terrestrial-mass planets that might survive, at semi-major axes of a few hundred AU.  These bodies are much smaller and closer to the Sun than the hypothetical Planet IX proposed by \citet{Batygin16} to account for asymmetries in the orbital distribution of distant KBOs.

\section{Initial Conditions}
\label{initCond}

\noindent
We report on ten sets of N-body simulations containing the Sun and the four giant planets, as well as several embryos (0.05--2 $M_\earth$) initialized on nearly circular and co-planar orbits within the orbits of the giant planets.  The inner planets are ignored.  The giant planets are initialized at their current semi-major axes, but with smaller inclinations and eccentricities (see below for details).  The parameters of these simulations are summarized in the first five columns of Table 1.  In each case, we represented the embryos by $N_{\rm eb}$ massive extra bodies (MEBs) of mass $M_{\rm eb}$ placed on orbits between Jupiter and Neptune.  The semi-major axes of the MEBs satisfy the relation 
\begin{equation}
a_{i+1} = a_{i} + \Delta a_i^{0.5}, i = 0, ... 
\end{equation} 
where $a_i$ is the semi-major axis of the $i^{\rm th}$ extra body.  If $a_{i+1}$ lies within two Hill radii of any of the four giant planets, we assume the orbit is unstable and we do not put a planet there.  We choose $a_0 = 5.91$ AU (two Hill radii outside Jupiter's orbit) and choose the largest value of $\Delta$ such that the most distant body would still be within the orbit of Neptune (30 AU) given that there are $N_{\rm eb}$ total bodies.  Apart from the gaps at the giant planets, this spacing corresponds to a surface density proportional to $R^{-1.5}$ as in Hayashi's (1981) minimum-mass solar nebula.  We ran 30 simulations in each set (except for a set of 10 simulations where we set the MEB mass to zero), differing only in the random number seeds.  
\par
Each simulation also contains a set of 50 test particles on orbits with semi-major axes 5 to 50 AU, with radial distribution corresponding to a $R^{-1.5}$ surface density profile. 
\par
The initial inclinations and eccentricities of all bodies, including the giant planets, are drawn from Rayleigh distributions with $\sigma_i =  0.001$ radians (= $0.057^\circ$) and $\sigma_e = 0.002$, having probability density  
\begin{equation}
\rho(i) = \frac{i}{\sigma_i^2} \exp{\left(-\frac{i^2}{2\sigma_i^2}\right)}; \quad \quad \rho(e) = \frac{e}{\sigma_e^2} \exp{\left(-\frac{e^2}{2\sigma_e^2}\right)}.
\label{eq:Rayleigh}
\end{equation}
\par
We integrated the orbits for using the IAS15 package, developed by \citet{Rein12}.  Within this package, we used the IAS15 integrator \citep{Rein15}, which is designed to handle close encounters and highly eccentric orbits.

We did not include the effects of Galactic tides, tides from the Sun's birth cluster, or passing stars.  We neglect Galactic tides and passing stars because we expect these to not have a significant effect on the few hundred AU scales of interest.  For example, the torque from the Galactic tide would need  $\sim 10^{12}$ years to impart enough angular momentum to convert a radial orbit to a circular one at 200 AU.  We do not include cluster tides even though they may well have affected the evolution of the outer solar system \citep{FB00}, as their magnitude and the time over which they act is uncertain, and in any case the goal of this paper is to focus on the effects of mutual interactions between the MEBs rather than the effects of cluster tides.

\begin{sidewaystable}[ht]
\centering
{\small
\vspace{5mm}
\caption{Simulation Statistics}
\begin{tabular}{c c c c c c c c c c c c c c}
\\
\hline\hline
1 & 2 & 3& 4& 5& 6& 7 &8 &9 &10&11&12&13 \\
 Sim set & N$_{\rm sims}$ & $N_{\rm eb}$ & $M_{\rm eb}$($M_\earth$) & $M_{\rm eff} (M_\earth)$  & $f_s$& $\langle r \rangle$  &$\langle N_{\rm remain} \rangle$& $\langle f \rangle$ & DD with & DD no &Kuiper&Kuiper   \\
 & & & & & & & & & MEB & MEB & Belt & RMS Inc. \\ [0.5ex] 
\hline 
1 & 30 & 10 & 2.0  &6.3&0.67  &4.5\%   &0.5  &0.19  & 0.5\%      &0.5\%   & $0.004\%$ & N/A \\
2 & 30 & 20 & 1.0  &4.5&0.87&2.5\%   &0.5    &0.26  & 0.5\%   & 2.0\%  & $0.05\%$  & 29$^\circ$ \\
3 & 30 & 40 & 0.5  &3.2&  0.97&3.2\%   &1.3    &0.18  &1.8\%    &0.9\%   & $0.09\%$ & 23$^\circ$\\
4 & 30 & 10 & 1.0  &3.2&0.80&3.3\%   &0.3  &0.17  & 1.0\%  &2.5\%   &$0.22\%$ & 29$^\circ$ \\
5 & 30 & 20 & 0.5  &2.2&  1.0&3.3\%   &0.7    &0.13   & 1.9\%  &0.8\%   & $0.14\%$ & 23$^\circ$ \\
6 & 30 & 40 & 0.25&1.6&  1.0&2.4\%   &1.0    &0.12     & 1.0\%  &2.0\%   & $0.12\%$ & 25$^\circ$ \\
7 & 30 & 20 & 0.25&1.1&  1.0 & 1.2\%  &0.2 & 0.27 & 1.4\%  &0.3\%   &$0.42\%$ & 19$^\circ$ \\
8 & 30 & 40 & 0.1  &0.6&  1.0 & 1.1\%  &0.4    & 0.22   & 0.2\%  &0.4\%   & $1.4\%$ & 12$^\circ$ \\
9 & 30 & 40 & 0.05&0.3&  1.0 & 1.6\%  & 0.6&0.23  & 0.0\%  &0.0\%   & $4.3\%$ & 7$^\circ$ \\
10 & 10 & 40 & 0.0&0.0&  1.0 & 34.0\%&13.6 & 1      & 0.0\%  &N/A       & $21\%$ & 0.8$^\circ$ \\[1ex]
\hline
\end{tabular}
\raggedright
\\
\vspace{3mm}
Column 2: Number of simulations in this set.\\
Column 3: Number of massive extra bodies (MEBs). \\ 
Column 4: Mass of each MEB. \\
Column 5: $M_{\rm eff}  \equiv M_{\rm eb} \sqrt{N_{\rm eb}}$ is proportional to the magnitude of the torques experienced by the MEBs if they are randomly distributed. \\
Column 6: Fraction of systems in which all four giant planets remained for the duration of the simulation. \\
Column 7: Mean fraction of MEBs that remain in the simulations included in column 6.\\
Column 8: Mean number of MEBs that remain in the simulations included in column 6 (column 3 times column 7). \\
Column 9: Fraction of the systems with at least one remaining MEB, in which all MEBs would be expected to escape dynamical detection and be fainter than a $V$-band magnitude of 19 (see Section \ref{observationalConstraints} for details).\\
Column 10:  Fraction of test particles that end up in the detached disk ($q > 38$ AU and $80 {\rm \, AU} < a < 500 {\rm \, AU}$) for simulations with at least one remaining MEB.\\
Column 11:  Fraction of test particles that end up in the detached disk ($q > 38$ AU and $80 {\rm \, AU} < a < 500 {\rm \, AU}$) for simulations with no remaining MEBs.\\
Column 12: Fraction of test particles that end up in the Kuiper belt (30 AU $<R<$ 50 AU). \\
Column 13: RMS inclination of particles in Kuiper belt (30 AU $<R<$ 50 AU). \\
}
\label{table:initialConditions}
\end{sidewaystable}
\clearpage

By starting the outer planets at their current semi-major axes we have neglected migration of their orbits due to gravitational interactions with the gaseous protoplanetary disk or planetesimal disk.  Neglecting the gaseous disk is justified because it probably disappeared within a few Myr of the formation of the solar system, whereas the characteristic evolution timescale in our simulations is much longer, $\sim 100$ Myr (see Figure \ref{timePlot}). Neglecting the planetesimal disk is more serious, because there is strong indirect evidence for migration, such as the 3:2 mean-motion resonance between Neptune and Pluto \citep{Malhotra93}. However, the migration history of the outer planets is uncertain, and given the preliminary nature and computational expense of this investigation we did not investigate a variety of migration models. 
\par
We removed bodies from the simulation if they went outside a box with side length $5\cdot 10^3$ AU centered on the Sun.  Outside this box, our neglect of the Galactic tides is not accurate.  If two particles collide, they are assumed to merge completely.  In determining whether a collision occurred, we used the current radii of the four giant planets. We also removed any particle that came within 0.1 AU of the Sun.  Integrations were run for 4.5 Gyr, or until all of the MEBs and test particles, or one of the giant planets, were removed from the simulation.  
\par
Simulation sets 1--3 and 4--6 have total masses of $20\,M_\earth$ and $10\,M_\earth$ respectively in MEBs, with varying masses for the individual MEBs between 0.25 and 2 $M_\earth$.  Simulation sets 7--9 explore smaller total masses of MEBs.  Set 10 is a control set where the mass of the MEBs is set to zero.  We also ran more simulations to explore parameter space; these informed our conclusions but will not be reported on explicitly.

\section{Simulation Results}
\label{simpleModel}

\noindent
When reporting orbital elements, we use Jacobi coordinates.  Inclinations are reported relative to the fixed reference plane near which the bodies were initialized.
 \begin{figure}[htp]
\centering
\includegraphics[width=1.0\textwidth]{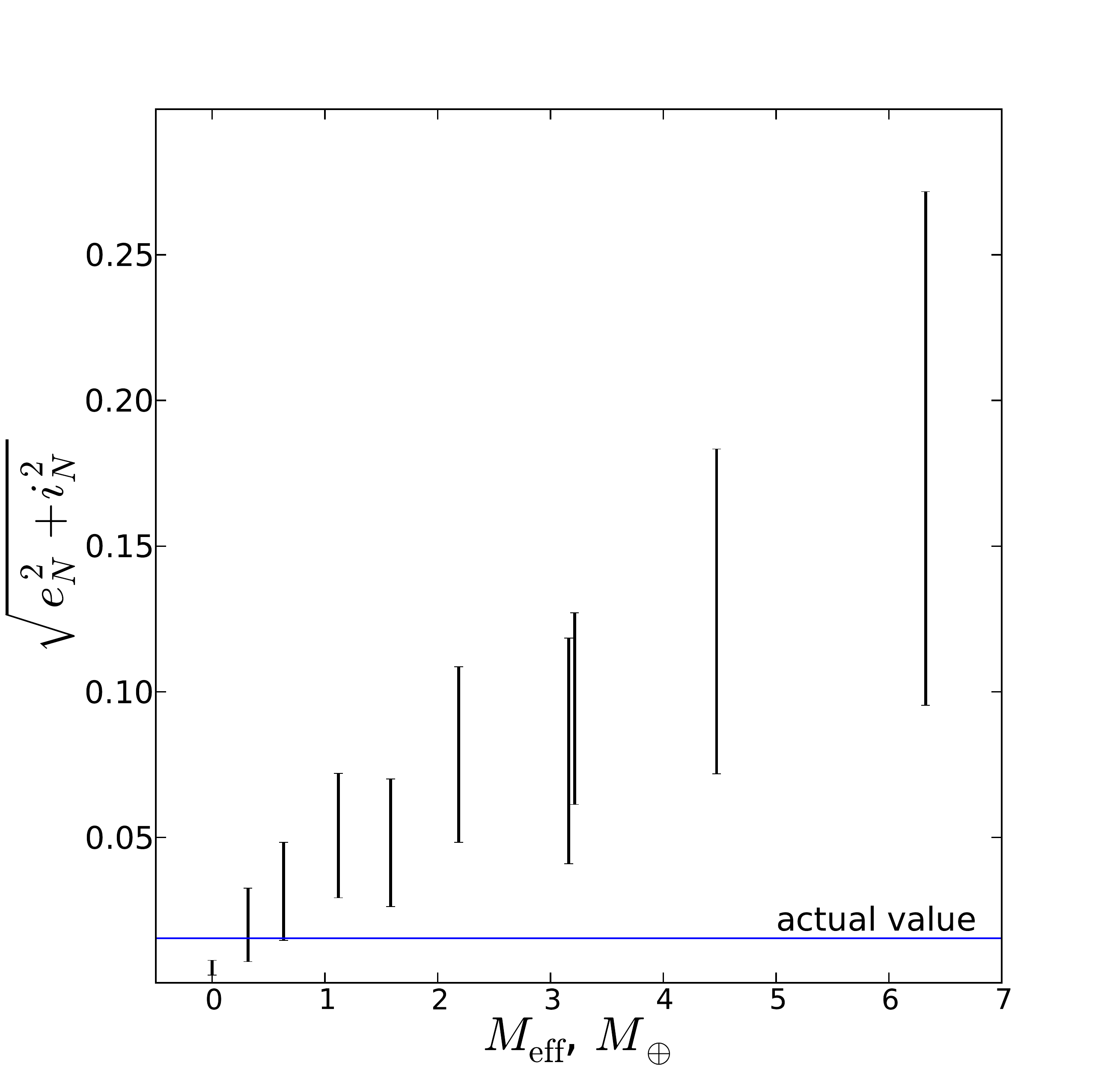}
\caption{Excitation of Neptune's eccentricity and inclination ($\sqrt{e_N^2+i_N^2}$) vs. $M_{\rm eff} = M_{\rm eb} \sqrt{N_{\rm eb}}$.   We have plotted the interdecile range for all surviving simulations (simulations in which all 4 giant planets are still present at the end) in each set.  The horizontal line corresponds to the averaged (over secular oscillations) present-day value of $\sqrt{e_N^2+i_N^2}$, where $i_N$ is measured relative to the invariable plane of the solar system.  The two lines near $M_{\rm eff}= 3.1M_\earth$ have been horizontally offset slightly so as to both be visible.  $M_{\rm eff}$, which is proportional to the RMS torque from the massive extra bodies (MEBs) if they are randomly distributed, is a good predictor of the typical value of $\sqrt{e_N^2+i_N^2}$.   }
\label{excitation1}
\vspace{-.05cm}
\end{figure} 

 \begin{figure}[htp]
\centering
\includegraphics[width=1.0\textwidth]{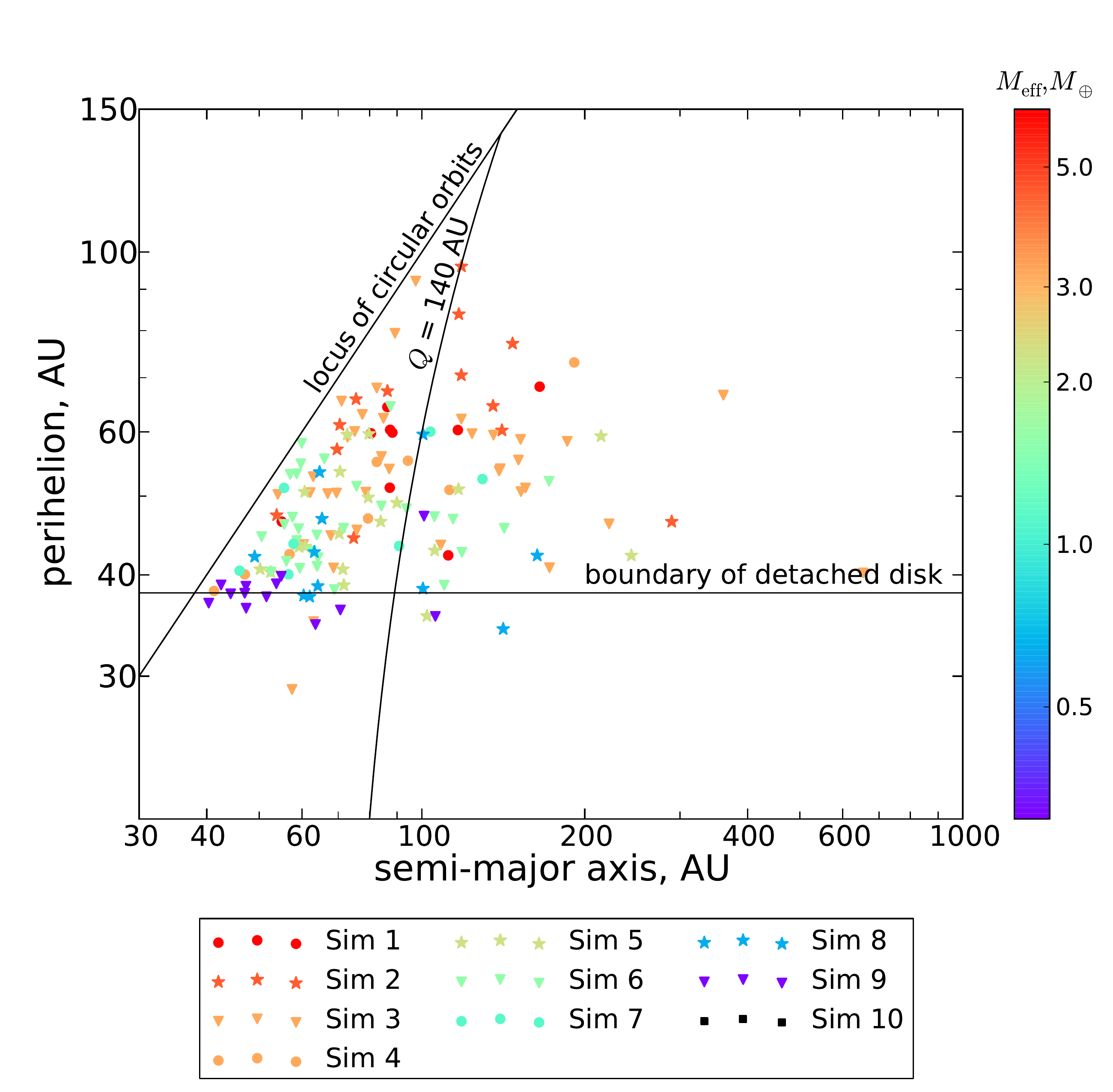}
\caption{Location of massive extra bodies (MEBs) in the space of semi-major axis and perihelion at the end of the simulations.  The color of the symbols corresponds to the effective mass of the population of MEBs.  The black squares correspond to the case in which the MEBs have zero mass.  The legend refers to the simulation numbers in column 1 of Table 1.  We only show bodies in surviving systems (systems in which all four giant planets remain).     We have labeled the lines corresponding to circular orbits, the boundary of the detached disk, and to orbits with aphelion equal to 140 AU (a rough bound to detectability, see Section \ref{observationalConstraints}).  While many bodies from simulation set 10 ($M_{\rm eb} = 0$) survive, their orbits have semi-major axes less than 30 AU, and are not shown on this plot.}
\label{AQone}
\vspace{-.05cm}
\end{figure}

 \begin{figure}[htp]
\centering
\includegraphics[width=1.0\textwidth]{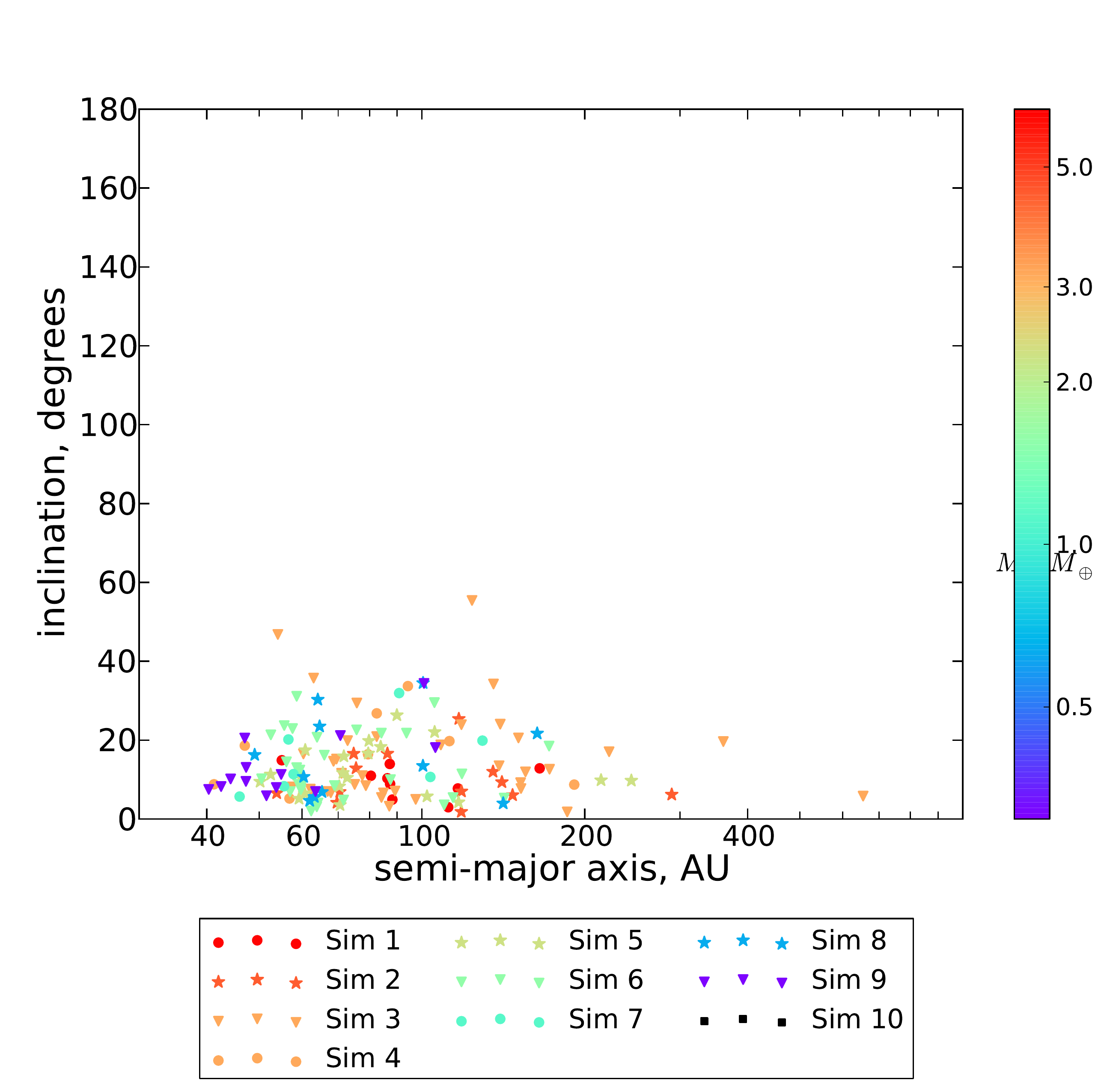}
\caption{Location of MEBs in the space of semi-major axis and inclination at the end of the simulations.  We have used the same color coding and point styles as in Figure \ref{AQone}.  Inclinations remain moderate and do not vary greatly with semi-major axis or $M_{\rm eff}$. }
\label{AIone}
\vspace{-.05cm}
\end{figure} 

 \begin{figure}[htp]
\centering
\includegraphics[width=0.9\textwidth]{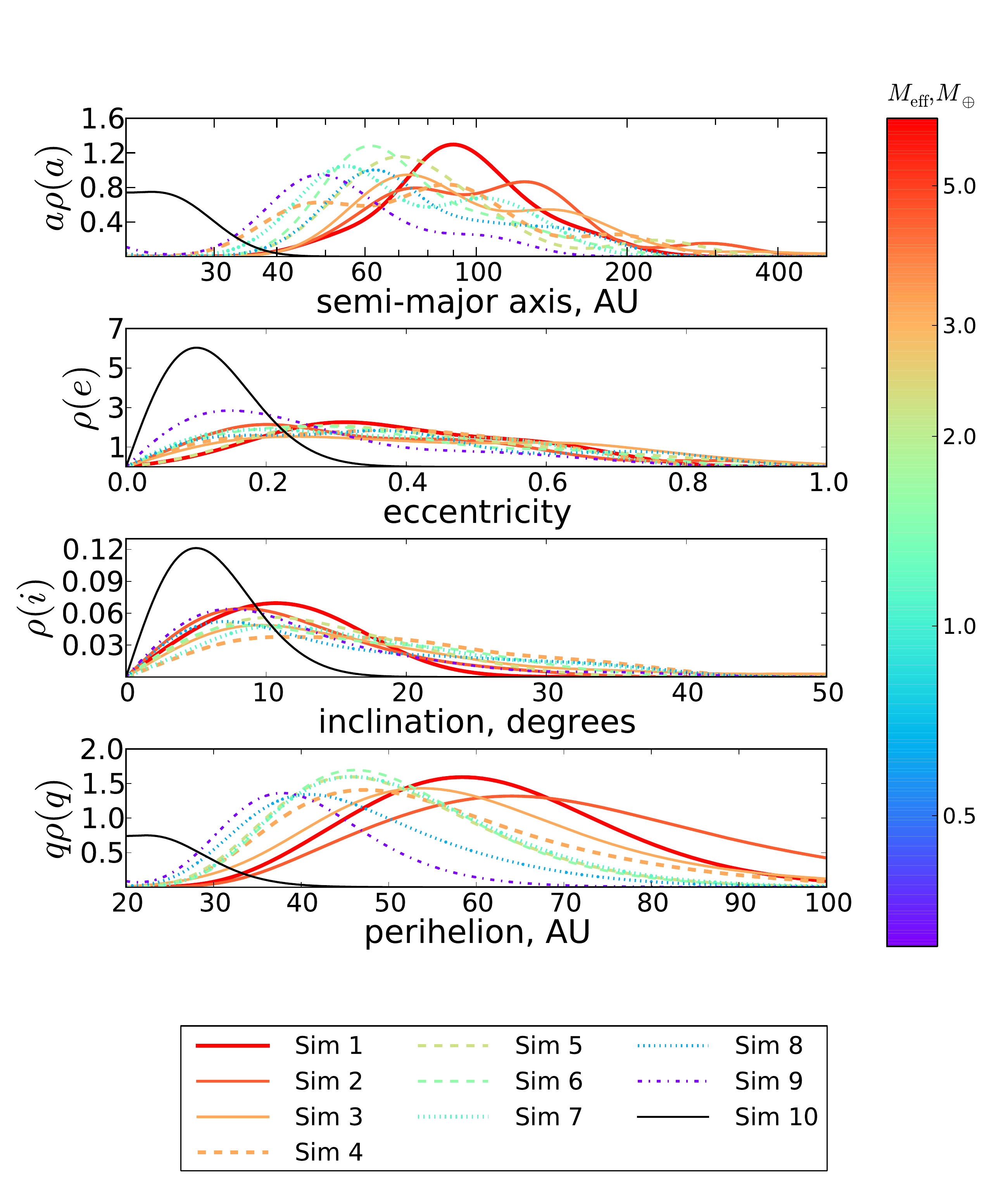}
\caption{Smoothed estimates of the density of the MEB population in semi-major axis, eccentricity, inclination and perihelion.  The line color corresponds to $M_{\rm eff}$, and the legend refers to the simulation numbers in column 1 of Table 1.  The drop of all the curves $\rho(e)$, $\rho(i)$ to zero at $e = 0$ and $i = 0$ is due to our smoothing kernel and is expected for any smooth distribution in phase space (see text).  The semi-major axis distributions peak between 50 and 150 AU.  Higher values of $M_{\rm eff}$ lead to higher mean values of the perihelion $q$.}
\label{rhoaei}
\vspace{-.05cm}
\end{figure} 

 \begin{figure}[htp]
\centering
\includegraphics[width=1.0\textwidth]{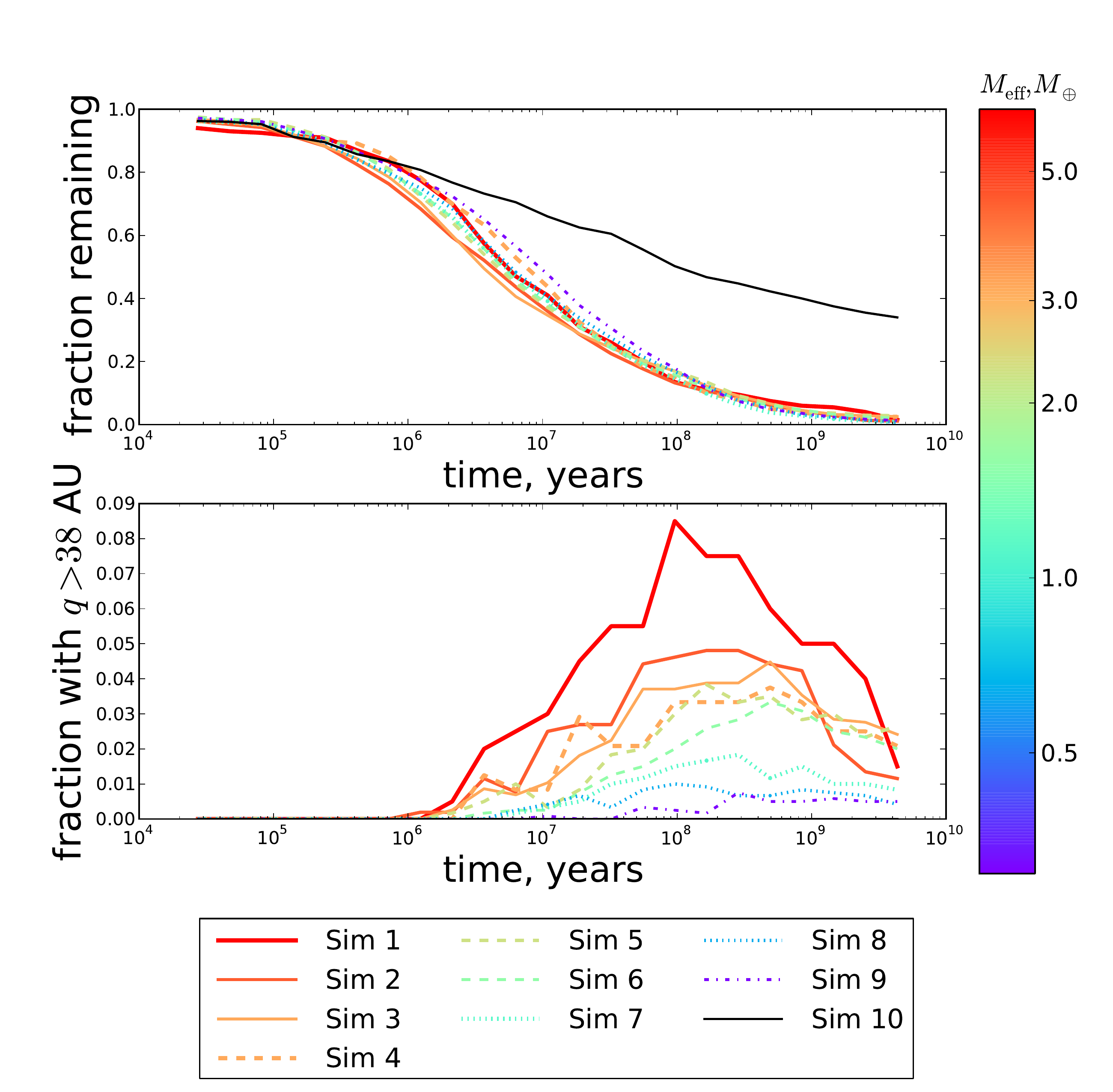}
\caption{The top panel shows the fraction of remaining MEBs as a function of time.  The bottom panel shows the mean fraction of MEBs in the detached disk (perihelion greater than $38$ AU and aphelion in the range 80 AU $\leq a \leq$ 500 AU) as a function of time.  The detached disk forms over a few times $10^8$ years, and decays slowly afterwards.   }
\label{timePlot}
\vspace{-.05cm}
\end{figure} 

Figure \ref{excitation1} shows the interdecile range (10$^{\rm th}$ to 90$^{\rm th}$ percentile) of $\sqrt{e_N^2 + i_N^2}$ (where $e_N$ and $i_N$ are the eccentricity and inclination of Neptune) as a function of the effective mass of the MEB population, given by 
 \begin{equation}
 M_{\rm eff}  = M_{\rm eb} \sqrt{N_{\rm eb}}.
 \end{equation}
If the MEBs are randomly distributed azimuthally, then $M_{\rm eff}$ is proportional to the net torque that the MEBs exert on each other and smaller planetesimals.  The figure shows that $M_{\rm eff}$ is a good predictor of the eccentricity and inclination excitation of Neptune.  Similar results apply to Uranus.

All but the smallest values of $M_{\rm eff}$ considered in our simulations excite the eccentricity and inclination of Neptune to values that exceed those in the real solar system.  It seems we require $M_{\rm eff} \lesssim M_\earth$ in order for the eccentricity and inclination of Neptune to be roughly compatible with the current observations.  The constraints from Uranus are much weaker due to its higher current eccentricity of 0.046.  Only simulations 8 and 9 have the current value of $\sqrt{e_N^2 + i_N^2}$ within their interdecile range, but the current value of $\sqrt{e_N^2 + i_N^2}$ is not far outside the range for simulations 6, 7 and 10.  Subsequent damping of the eccentricities and inclinations (see Section \ref{damping}) could relax this limit.

Column 6 of Table 1 shows the fraction of systems that survive in each simulation set.  We define a simulation as ``surviving" if all four giant planets remain on bound orbits at the end of the simulation.  A significant fraction of simulations in the sets with the highest values of $M_{\rm eff}$ did not survive.  Since Figure \ref{excitation1} indicates that the degree of eccentricity and inclination excitation is roughly proportional to $M_{\rm eff}$, we conclude that if $M_{\rm eff} \gtrsim 5 M_\earth$ there is a significant chance that one of the giant planets will be ejected or collide with the Sun or another planet (unless the eccentricities and inclinations are subsequently damped, see Section \ref{damping}).  In the figures and discussion that follow, we only report results from surviving simulations.  
\par
Figure \ref{AQone} is a scatter plot of perihelion vs. semi-major axis for MEBs remaining in the solar system.  Points are color-coded according to the value of $M_{\rm eff}$ for the corresponding simulation.  The black squares correspond to our control simulation with massless extra bodies (simulation 10).  We have labeled three special lines.  The line at  $q = a$ corresponds to circular orbits.  The horizontal line labeled ``boundary of detached disk" at $q = 38$ AU is roughly the distance outside which non-resonant perturbations from Neptune are unimportant over the age of the solar system \citep{Gladman02}.  It is expected that after several Gyr, there will be few non-resonant bodies inside this line, as most would have been ejected by Neptune.  Finally, we have drawn a line where the aphelion $Q$ is equal to 140 AU.  This is a rough lower limit to the distance at which an MEB of $1M_\oplus$ could have escaped detection (see Section \ref{observationalConstraints}).

We see that most of the remaining MEBs are on orbits with perihelia less than 80 AU and semi-major axes less than a few hundred AU.  In contrast, the test particles in the model of \citet{Brasser12}, where the torque was applied by tides and stellar flybys in the birth cluster, typically have semi-major axes greater than 1,000 AU.
\par
Figure \ref{AIone} is a scatter plot of orbital inclination against semi-major axis.  There are no remaining MEBs with inclination greater than 60$^\circ$.  The median inclinations of the surviving MEBs for our different simulation sets (not including set 10) range from $7^\circ$ to $18^\circ$, although much of this scatter is due to small number statistics.  This is very different from the inclination distribution of detached objects that are formed in models involving cluster tides and flybys \citep{Brasser12}---even in the innermost part of their cloud, the median inclination is between 45$^\circ$ and 55$^\circ$.  Thus, the inclination distribution in the detached disk provides a clean test to distinguish our model from cluster tides.
\par
Figure \ref{rhoaei} shows the density of the MEBs in semi-major axis, eccentricity, inclination and perihelion.  The curves for semi-major axis and perihelion were smoothed with a log-normal kernel, with a dispersion of 0.2 in $\ln{a}$ or $\ln{q}$.  To determine the curves for eccentricity, we smoothed the eccentricity vector with a 2D Gaussian with $\sigma_e = 0.1$, and then integrated the resulting probability distribution over angle to recover the distribution of scalar eccentricity.  This procedure ensures that $\rho(e) \sim e$ for small values of $e$---consistent with a non-singular distribution in phase space around $e = 0$.  We applied the same method to the inclination using $\sigma_i = 5^\circ$.  Median eccentricities range from 0.19 to 0.40 (excluding simulation set 10), although as can be seen from the figure, the distributions are quite broad.  Values of eccentricity near unity are uncommon because particles are mostly limited to $0 < e < 1 - q_{\rm crit}/a$, where $q_{\rm crit} \approx 38$ AU is the perihelion below which a planet will be ejected by perturbations from Neptune.  Systems with higher values of $M_{\rm eff}$ have higher mean values of the perihelion.  This can be understood qualitatively as being due to the larger typical torques experienced by particles in these simulations.

Figure \ref{timePlot} shows the evolution of the number of MEBs over time.  The top panel shows the total fraction of remaining MEBs.  Most of them are removed over $\sim 10^7$ year timescales.  In contrast, about a third of the MEBs survive for the duration of the simulation if their masses are negligible (black line, simulation set 10).  This difference arises because the stability of test particles on near-circular orbits in the region of the giant planets (5--30 AU) depends strongly on the eccentricities of the planets. In our simulations the planets start on nearly circular orbits (Equation 2) but MEBs with non-zero masses rapidly excite the eccentricities, while those with zero masses do not.  The bottom panel shows the fraction of MEBs on orbits with $q > 38$ AU.  This number grows for a few times $10^8$ years, and slowly declines thereafter.  As expected, there are no bodies with $q > 38$ AU if the MEBs have zero mass (i.e., there is no black line in the bottom panel of Figure \ref{timePlot}).  The approximate timescale for the creation of bodies with perihelion greater than 38 AU can be derived simply.  Assuming a typical specific torque of $GM_{\rm eff}/\langle R \rangle$, where $\langle R \rangle$ is the average orbital separation, then the timescale to produce orbits with perihelion greater than $q_{\rm crit}$ can be approximated as 
\begin{equation}
\label{formationTime}
\tau = \frac{\sqrt{2GM_\odot q_{\rm crit}}}{GM_{\rm eff}/\langle R \rangle } = 7 \cdot 10^7 \left(\frac{M_{\rm eff}}{M_\earth}\right)^{-1}\frac{\langle R \rangle}{150\, \rm AU} \, \rm{years},
\end{equation}
where we have used $q_{\rm crit} = 38$ AU in the numerical estimate.  Equation \eqref{formationTime} agrees fairly well with the results shown in Figure \ref{timePlot}.  In particular, higher values of $M_{\rm eff}$ lead to more rapid emplacement of bodies in large-perihelion orbits.  On the other hand, after a group of MEBs is placed on large-perihelion orbits, higher values of $M_{\rm eff}$ also lead to more rapid depletion over the remaining age of the solar system.  Because the MEB population evolves on such long timescales, we expect the results of this model to not depend strongly on the details of the initial conditions and early evolution of the solar system.
\par
\begin{figure}[htp]
\centering
\includegraphics[width=0.8\textwidth]{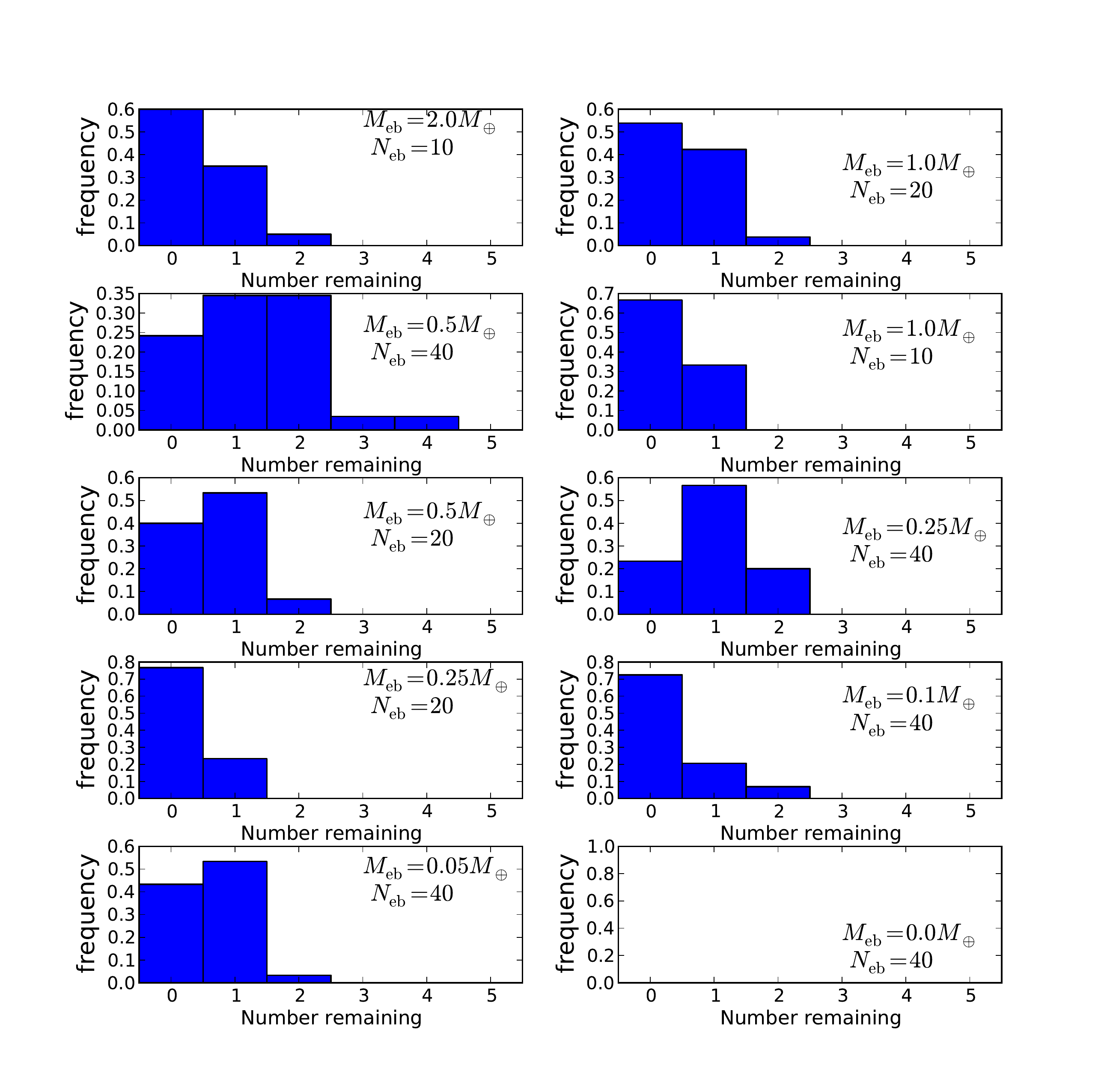}
\caption{Histograms of the number of MEBs present with semi-major axes greater than 38 AU at the end of the surviving simulations.  The mass and initial number of the extra bodies are labeled on each panel.  Simulations with intermediate-mass MEBs (between 0.25 and 0.5 $M_\earth$) tend to retain a higher number than simulations with either larger or smaller masses.  It is rare for more than two MEBs to remain.  }
\label{MPone}
\vspace{-.05cm}
\end{figure}

Figure \ref{MPone} shows a histogram of the number of remaining MEBs with $a > 38$ AU in the different runs of each simulation set.  A larger fraction of the MEBs remains to the end of the simulations in systems with intermediate-mass MEBs. There are few systems with more than one remaining MEB in which $M_{\rm eb}$ is greater than 0.5 $M_\earth$, presumably because the mutual torques between MEBs of larger masses are sufficiently strong to lower the perihelia to the point that MEBs are ejected by one of the giant planets, until only one MEB is left. 
\par
To summarize, our simulations show that a population of MEBs with effective mass $M_{\rm eff}=\sqrt{N_{\rm eb}}M_{\rm eb}$ exceeding $\sim 1M_\oplus$ will excite the eccentricities and inclinations of Uranus and Neptune to values larger than observed---thus either there has been subsequent damping, or $M_{\rm eff}\lesssim M_\oplus$. In the latter case, for $M_{\rm eb}\gtrsim 0.05M_\oplus$ we find a probability of up to $\sim 50\%$ that one or occasionally more MEBs survive in bound orbits until the present time. Their orbits typically have perihelia of 40--70 AU, semi-major axes less than 200 AU, and inclinations $\lesssim30^\circ$.

\par

\section{Fate of the Test Particle Population}

\noindent
As described in Section \ref{initCond}, in each set of simulations we also included 50 test particles that were originally in dynamically cold orbits (initial eccentricity and inclination less than 1\%) between 5 and 50 AU, with surface density $\propto R^{-1.5}$.  These are included to monitor the effect of the MEBs on the Kuiper belt, and to study the efficiency of injection of particles into the detached disk (defined in this paper as containing any body with perihelion greater than 38 AU and semi-major axis between 80 and 500 AU) and the Oort comet cloud.  \citet{Gladman06} showed that massive bodies exterior to Neptune can lift test particles to large perihelion distances, thus providing a natural mechanism to create a disk of detached objects such as Sedna \citep{Brown04} or 2012VP$_{113}$ \citep{Trujillo14}.  We make predictions for the the mass and orbital distribution of a detached disk produced in this way. Many of our results are similar to those of \citet{Gladman06}.

The Kuiper belt exhibits a rich dynamical structure. In addition to the detached disk there are Plutinos in the 3:2 resonance with Neptune, objects in other mean-motion resonances (1:2, 3:5, 4:7, etc.), a ``scattered disk'' consisting of objects with perihelion $q\lesssim 38$ AU and eccentricity $e\gtrsim 0.2$, and ``classical'' Kuiper belt objects (KBOs), which have low eccentricities and semi-major axes in the range 40--50 AU. The classical population is generally divided into ``cold'' and ``hot'' components, with RMS inclinations of about $3^\circ$ and 10--$20^\circ$, each containing about half of the total population  \citep{Brown01,Gulbis+10}.  Because our simulations have far fewer test particles than the number of known KBOs we cannot compare our model predictions to the detailed structure of the Kuiper belt. Thus we restrict ourselves to a few brief, crude comparisons.

\begin{figure}[htp]
\centering
\includegraphics[width=0.9\textwidth]{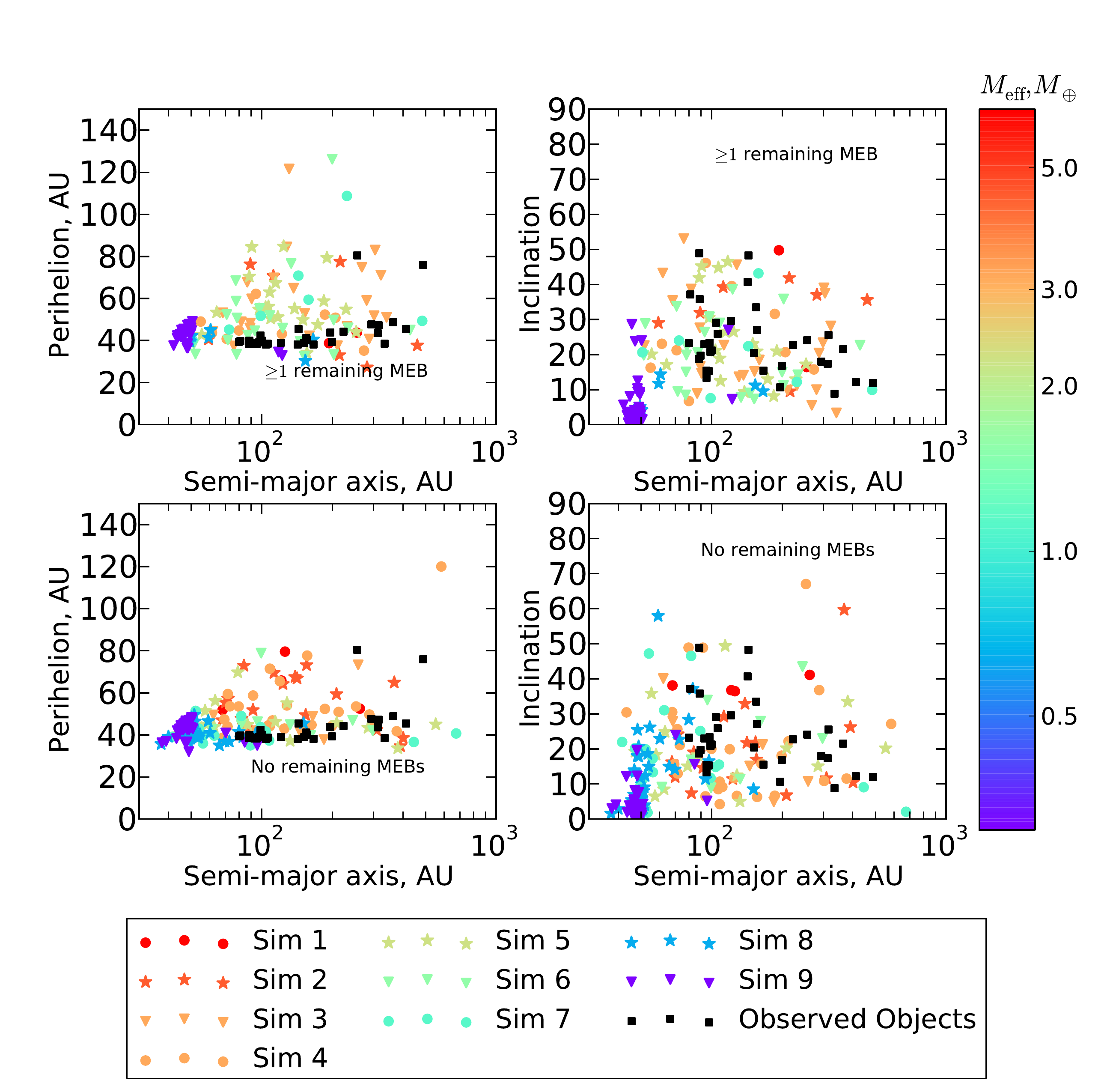}
\caption{Orbital properties of the test particle population.  The black squares represent known members of the detached disk ($q > 38$ AU, 80 AU $<a<$ 500 AU);  we did not plot observed objects with $a < 80$ AU as they are quite numerous and do not fall under our definition of detached disk objects.   Otherwise, the point properties described in the legend are the same as for Figure \ref{AQone}, except that simulation 10 is not plotted.  The top panels are for surviving runs in which at least one of the MEBs remains at the end of the simulation.  The bottom panels are for surviving runs in which all of the MEBs were ejected.  There is an observational selection bias that favors the discovery of objects with small semi-major axes and perihelia. }
\label{KuiperPoster}
\vspace{-.05cm}
\end{figure}

 Figure \ref{KuiperPoster} shows scatter plots of perihelion and inclination vs.\ semi-major axis for all test particles still present at the end of our simulations.  We divide the results into top and bottom panels based on whether any of the MEBs are still present at the end of the simulation, to see if there are clues in the distributions as to whether or not the solar system still has an undiscovered MEB.  As expected, almost all of the test particles with perihelion $\lesssim 38$ AU have been ejected. The orbits of the remaining test particles are similar to those of the remaining MEBs  (Fig.\ \ref{AQone}), and seem not to depend strongly on whether any of the MEBs remain.  The inclination distribution of the test particles is slightly broader than that of the MEBs.  The inclination and semi-major axis distributions are consistent with observations of the detached disk.  Our results also imply that few bodies will be found in the detached disk with semi-major axes $\gtrsim 500$ AU.

Column 10 in Table 1 shows the fraction of test particles that end up in the detached disk for simulations in which at least one MEB remains.  Column 11 shows this same fraction, but for simulations in which no MEBs remain.  We find typically that around 1\% of the initial population is transferred to the detached disk, with a variation of a factor of two or so among the simulations. 

For comparison, the efficiency of transferring planetesimals to the Oort cloud is a few per cent \citep{KQ08,Brasser10,Dones+15,Shannon+15}, with the precise number depending on the initial semi-major axis distribution of the planetesimals,  the migration history of the outer planets, the definition of the minimum semi-major axis of the Oort cloud, and the history of the Sun's environment. In our model, the transfer efficiency to the detached disk is about 1\%, so we predict that the detached disk should be a factor of a few less massive than the Oort cloud. The mass of the Oort cloud is estimated to be $\sim 3M_\oplus$, with large uncertainties \citep[e.g.,][]{Morbidelli05}, so we arrive at an estimate of $\sim 1M_\oplus$ for the scattered disk. This estimate may be unrealistically large because of the well-known problem that the estimates of Oort cloud formation efficiency and mass used in this estimate also imply an unrealistically massive initial planetesimal disk \citep[e.g.,][]{Morbidelli04}.

We may also directly estimate the mass in the detached disk predicted by our model. The test particles in our simulations are distributed with surface density proportional to $R^{-1.5}$ from 5 to 50 AU \citep{Hayashi81}.  Following \cite{hahn99} and \cite{gomes04}, we assume that about $40M_\oplus$ of solid planetesimals were left in the disk between 20 and 50 AU after formation of the planets (larger masses cause Neptune to migrate rapidly to the outer edge of the disk, rather than stopping at its current position). With this assumption the total mass of our test-particle disk is $\sim 70M_\oplus$.  If 1\% of the test particles are transferred to the detached disk, then the detached disk should contain $\sim 0.5$--$1\,M_\oplus$. For comparison, \citet{Brown04} make a rough estimate that there is $\sim 5M_\oplus$ in bodies with Sedna-like orbits.  This estimate assumed that the Sedna-like population was isotropically distributed, so for the flattened inclination distribution observed in the detached disk a better estimate is probably $\sim 1M_\oplus$. This result was based on only one body (Sedna), so it is subject to considerable uncertainty.  \citet{Gladman02} estimate that there are $10^6$ bodies with diameter greater than 100 km in the detached disk.  This corresponds to a minimum mass of $0.2 M_\earth$, assuming a density of 2 g cm$^{-3}$.  For a plausible size distribution the actual mass could be larger by a factor of five or more. We conclude that the mass of the detached disk estimated by our model is consistent with the mass estimated by observers, although with substantial uncertainties in both approaches.

\citet{MB97} and \citet{Petit99} suggest that gravitational interactions with large planetesimals scattered outward by the giant planets could explain why the Kuiper belt is much less massive and more excited that it is thought to have been primordially.  This suggestion is qualitatively consistent with our finding that the MEBs remove most of the test particles in our simulations.  We now make a more quantitative comparison.  \citet{gladman01} estimate a current mass of $0.1M_\earth$ for the Kuiper belt, which they define to include all bodies (except for Neptune) in the distance range 30--50 AU from the Sun.  If we assume, as we did above, that the planetesimal population between 5 and 50 AU had a mass of 70 $M_\earth$, then about 0.14\% of this population is found in today's Kuiper belt. To compare this result to our simulations, we added the fractional times that each surviving test particle spent between 30 and 50 AU, and divided this sum by the total initial number of test particles. The results are shown in Column 12 of Table 1.  By this metric, it appears that simulations 2--7 do a reasonable job of reproducing the $0.14\%$ target\footnote{We also considered a more restricted definition of the Kuiper belt that only includes particles with semi-major axes between 30 and 60 AU, and weights them by the fraction of the time they spend between 30 and 50 AU.  This reduces contamination of the ``Kuiper belt" by objects in the scattered or detached disks.  This definition causes modest reductions in many of the Kuiper belt populations, but does not change the qualitative picture.  In this definition, fractions (0, $0.02\%$, $0.02\%$, $0.1\%$, $0.1\%$, $0.05\%$, $0.4\%$,  $1.3\%$, $4.3\%$, $21\%$) survive (cf. column 12 of Table 1) for simulations 1--10.  That said, in most cases we only have 1,500 test particles per simulation set, so results smaller than $\sim 0.5\%$ are uncertain because of small-number statistics.}.

An important concern is whether or not our model can reproduce the division of the classical Kuiper belt into cold and hot components. It is commonly argued \citep[e.g.,][]{Shannon+16} that the cold component formed {\it in situ} between 40--50 AU and has remained undisturbed since then, and this view is difficult to reconcile with a late stage of inclination excitation by MEBs. We note that the division into distinct cold and hot components is somewhat artificial, since it relies on the assumption that this inclination distribution of each component must have the Rayleigh form (\ref{eq:Rayleigh})\footnote{It is often said that the inclination distribution in the classical Kuiper belt is ``bimodal'' but this usage is misleading, as the observed distribution has only one statistically significant maximum.}. This assumption is often justified by the central limit theorem, but this theorem applies only if the inclination excitation occurs through a large number of weak perturbations. In contrast, if the excitation is dominated by a few strong perturbations the distribution will have fatter tails than a Gaussian. For example, \cite{collins07} show that the eccentricity distribution excited by shear-dominated encounters has the non-Rayleigh form $dN \propto e\,de(e_c^2+e^2)^{-3/2}$, where $e_c(t)$ is a time-dependent scaling factor. Similarly, the inclination distribution excited by MEBs may be dominated by a few close encounters and could perhaps produce the observed shape of the inclination distribution from a single population. Testing this hypothesis would require dedicated simulations in which the Kuiper belt is represented by a much larger test-particle population than we have in our runs. Existing simulations of the effects of MEBs on the Kuiper belt \citep{fer80,mprp15} are mostly for MEBs on low-eccentricity, low-inclination orbits and hence are not directly applicable. 
\par
Our simulations contain an insufficient number of remaining Kuiper belt particles for us to determine the functional form of the inclination distribution.  Column 13 of Table 1 shows the RMS inclinations of the remaining particles with 30 AU $<a<$ 60 AU, weighted by the time the particle spends interior to 60 AU.  Unless $M_{\rm eff} \lesssim M_\earth$ the particles have somewhat higher inclination than even the hot component in the parameterizations of the inclination distribution of the classical Kuiper belt given in \citet{Brown01,Gulbis+10}.

Related issues are (i) the presence of correlations in the classical Kuiper belt population between inclination and physical properties---color \citep{TR00,TB02,plj08,Terai+17}, size \citep{Fraser+14}, albedo \citep{Brucker09}, and binarity \citep{Noll+08}. These could reflect either different physical properties in the cold and hot populations, or inclination-dependent environmental effects on a single population; (ii) many KBOs are in mean-motion resonances with Neptune, and the number and mass of MEBs affects the fraction of resonant KBOs and their libration amplitudes \citep{NV16}. 

To summarize, our simulations show that MEBs can excite planetesimals into orbits with typical perihelia of 40--80 AU, semi-major axes as large as a few hundred AU, and inclinations 0--40$^\circ$, consistent with the observed properties and (within large uncertainties) total mass of the observed detached disk. The MEBs may also be responsible for destroying most of the Kuiper belt, which is much less massive than expected from extrapolating the solid content of the minimum solar nebula. These conclusions hold whether or not one or more MEBs is eventually found in the outer solar system, since the properties of the detached disk and Kuiper belt are mostly independent of whether or not any MEBs survive. Our simulations do not have enough test particles to determine whether interactions with the MEBs produce a residual Kuiper belt that is fully consistent with the observed dynamical structure of the Kuiper belt.

\section{Observational Constraints}
\label{observationalConstraints}

\noindent
MEBs can be detected dynamically or photometrically.  We first consider dynamical detection.  \citet{Fienga16} find that a hypothetical planet of $10M_\oplus$ would be dynamically detectable, mostly from its perturbations to Saturn, if it were currently closer to the Sun than about 370 AU.  This is a model-dependent limit, as they were considering a planet on a particular orbit proposed by \citet{Batygin16}, but probably the best one available for our purposes.  To good approximation a planet at that distance only interacts with the known planets via a stationary tidal potential over the interval of modern observations.  Thus we may assume that the detectability limit scales as $M/R^3$, where $M$ is the hypothetical planet's mass and $R$ its current distance.  Therefore an MEB should be dynamically detectable if its current heliocentric distance is less than about
\begin{equation}
\label{Rcriteq}
R_{\rm crit} = 170 \, \left(\frac{M}{M_\earth}\right)^{1/3} {\rm AU}.
\end{equation} 
\par
We next consider photometric detection.  The most extensive systematic survey is the Palomar Distant Solar System Survey \citep{Schwamb10}, which covered 30\% of the sky to a limiting $r$-band magnitude of 21.3.  Using serendipitous discoveries from the Catalina Sky Survey and the Siding Spring Survey, \citet{Brown15} estimate that there is a 30\% chance that the solar system contains an undetected KBO brighter than a $V$-band magnitude of 19.  In an abstract, \citet{Holman16} report that Pan-STARRS has completed a search for slow-moving objects brighter than an $r$-band magnitude of 22.5 over the entire sky north of $-30^\circ$ in declination (75\% of the sky).  This survey area includes most, but not all, of our bodies.  Assuming a geometric albedo of 0.04 and a constant density of 2.0 g cm$^{-3}$ (appropriate for long-period comets; Lamy et al. 2004), an $r$-band limit of 19.0 (22.5) corresponds to a critical detection distance of $140  \,(313) (M/M_\earth)^{1/6}$ AU.  This increases to $235\, (526) (M/M_\earth)^{1/6}$ AU if we take the albedo to be 0.32, appropriate for Sedna \citep{Pal12}.  Thus, the Pan-STARRs survey has a good chance of detecting MEBs, if any are still present. 

Column 9 in Table 1 shows the probability that none of the MEBs remaining at the end of a given simulation are either dynamically detectable or brighter than magnitude 19 assuming an albedo of 0.04.  The probability is only calculated for simulations in which at least one MEB remains.  To compute this probability, we use the smaller of $170 \, \left(M/M_{\earth}\right)^{1/3}$ AU (Equation \ref{Rcriteq}) and $140 (M/M_\earth)^{1/6}$ AU to calculate $R_{\rm crit}$ for each remaining MEB.  Then, given the orbital parameters of the MEB at the end of the simulation, we calculated the fraction of the orbital period that would be spent beyond $R_{\rm crit}$.  This is the probability that a given MEB would be undetected.  Then, to find the probability that all the planets are undetectable, we multiply all the individual probabilities.  This is not completely correct, even assuming the bodies' phases to be uncorrelated, as their dynamical effect on the known planets could add constructively or destructively depending on their current locations.  These probabilities are generally a few tens of percent, so if MEBs formed this way were still present in the solar system, it is likely but far from certain that they would have been detected.
\par
\citet{Volk17} find that there is a statistically significant (at the 97\% level) warp in the mean plane of the Kuiper belt by comparing the orbits of non-resonant bodies with semi-major axes in the range from 42 to 48 AU to those with semi-major axes from 50 to 80 AU.  They comment that this warp could be caused by an unseen Mars-mass body orbiting at 65--80 AU, consistent with the properties of the remaining MEBs in our simulations.

\section{Damping}
\label{damping}

\noindent
We found (Section \ref{simpleModel}) that the eccentricity and inclination of Neptune were excited above their observed values when $M_{\rm eff} \gtrsim M_\earth$.  This constraint on $M_{\rm eff}$ could be relaxed if the eccentricity and inclination were subsequently damped.  Multiple groups \citep[e.g.][]{Kokubo95,Tsiganis05} have found that the presence of many small bodies in a disk can damp the eccentricity of larger planetary-mass objects through dynamical friction.  Directly modeling this process in our simulations over the lifetime of the solar system was impractical, even ignoring interactions between planetesimals.  Instead, we experimented with implementing a drag force 
\begin{equation}
a_{r, i} = -\frac{v_{r, i}}{\tau_{\rm damp}} - \frac{v_{z, i}}{\tau_{\rm damp}},
\end{equation}
\\
where $a_{r, i}$, $v_{r, i}$ and $a_{v, i}$, $v_{z, i}$ are the radial and vertical accelerations and velocities of the $i^{\rm th}$ giant planet.  We assumed the damping time $\tau_{\rm damp}$ to be given by
\begin{equation}
\tau_{\rm damp}(t) = \tau_1 \exp{(t/\tau_2)}.
\end{equation}
We take $\tau_1 = 10^5$ years and $\tau_2 = 10^7$ years.  $\tau_2$ is consistent with the time over which small bodies are removed from the system (see Figure \ref{timePlot}).  The damping force does {\it not} act upon the MEBs, as they are presumed to be insufficiently massive to be affected by dynamical friction with a planetesimal disk.
\par
This formalism does produce some damping of the semi-major axes as well, but provided that the eccentricities and inclinations are much less than unity, one can show that under the influence of damping, 
\begin{equation}
\frac{d \ln{(a)}}{dt} = \frac{d}{dt}(e^2 + i^2).
\end{equation}
Therefore, if the eccentricities and inclinations remain low, they are damped much faster than the semi-major axis.
\par 
We ran simulations analogous to our simulation sets 1--3 using the damping scheme discussed above and found little qualitative difference from the results presented in this paper.  Even with $\tau_1 = 10^5$ years, large MEBs ($M_{\rm eb}\geq 2 M_\earth$) could still eject one or more of the giant planets.  As can be seen in the second panel of Figure \ref{timePlot}, the MEB population still interacts with the giant planets after more than $10^8$ years, at which point one would expect most of the planetesimal disk to be gone, and the damping to therefore be negligible. 

\section{Conclusions}

\noindent
It is highly likely that multiple ``planetary embryos" or massive extra bodies (MEBs) of up to a few Earth masses form among the giant planets.  These are scattered outwards by gravitational perturbations from the giant planets.  The MEBs exert torques on one another before they can be ejected from the solar system, thus increasing their perihelia beyond the gravitational reach of the giant planets.  They can therefore remain in the solar system for 4.5 Gyr,  outside the orbit of Neptune but far inside the Oort cloud.  Additionally, these bodies exert torques on smaller planetesimals, thereby naturally creating a detached disk containing objects with orbits similar to Sedna and  2012VP$_{113}$.  

The evolution of the population of MEBs is characterized by their effective mass $M_{\rm eff} = M_{\rm eb} \sqrt{N_{\rm eb}}$.  If $M_{\rm eff}$ exceeds about an Earth mass, then the eccentricity and inclination of Neptune are usually excited above the observed values, which rules out this range of parameter space unless some process subsequently damps them.
However, significant effects occur for smaller values of the effective mass.  For example, if 20 bodies of 0.25 Earth masses were present (Simulation 7; $M_{\rm eff} = 1.1 M_\earth$), then in 23\% of cases one MEB will remain in a moderate inclination orbit with semi-major axis between 45 and 130 AU, and on average 0.5\% of the mass in the planetesimal belt between 5 and 50 AU is transferred to the detached disk.  In many of our simulations with remaining extra bodies, at least one of them would be expected to have been detected either dynamically or photometrically, although this result depends strongly on studies done for other purposes or surveys that are still in progress.  In a few tens of percent of simulations with remaining MEBs, the bodies were remote enough to escape dynamical and photometric detection so far. 

The extra bodies that we discuss in this paper are much smaller and closer than the Planet IX proposed by \citet{Batygin16}: $\sim 0.1$--$0.5\,M_\oplus$ compared to 5--$20\,M_\oplus$ for Planet IX, and $\lesssim 200$ AU compared to 400--1000 AU for Planet IX. The motivation for our model is also different from the motivation for Planet IX: ours is motivated by simple considerations arising from the standard model for the formation of the outer planets, while the Planet IX hypothesis is motivated by asymmetries in the orbital element distribution of a selected set of KBOs. 

MEBs also transfer a few percent of the mass of the initial planetesimal belt into a detached disk composed of bodies on moderately inclined orbits with perihelia greater than 38 AU and semi-major axes between 80 and 500 AU.  Like the observed sample of 33 objects in our definition of the detached disk, the population of test particles in our simulations are on moderately inclined orbits.  The amount of material in the detached disk does not depend strongly on whether any of the MEBs remain in the system; thus the viability of this mechanism for producing the detached disk does not require the discovery of new terrestrial-mass planets in the outer solar system.

The original motivation for this model was to explain the properties of the detached disk, which is not predicted by the standard model of the formation and evolution of the Oort comet cloud. There are other tensions between observations and the standard model. These include a predicted mass for the Oort cloud that is too low given the likely mass in the planetesimal disk, a predicted mass for the scattered disk of comets that is too high, and a predicted inclination distribution for long-period comets with too many retrograde orbits (e.g., \citealt{Wiegert99, Morbidelli05}). It is possible that the presence of massive extra bodies in the outer planetesimal disk will alleviate or resolve some of these tensions, but studying this possibility is beyond the ambition of the current paper.
\par
We acknowledge helpful discussions with Matthew Holman, Brian Lacki, Matthew Payne, Roman Rafikov, and Yanqin Wu.  We also thank the anonymous referee for constructive criticism that significantly improved the paper.
\bibliographystyle{apj}
\bibliography{apj-jour,miniOortPaper_iteration_threeArxiv.bib}

 \end{document}